\documentclass{emulateapj}
\shorttitle{Eclipse depth-metallicity correlation}
\shortauthors{Dodson-Robinson}
\begin{document}

\title{A Correlation Between the Eclipse Depths of {\it Kepler}
Gas Giant Candidates and the Metallicities of their Parent Stars}

\author{Sarah E. Dodson-Robinson}

\affil{Astronomy Department, University of Texas at Austin, 1 University
Station C1400, Austin, TX 78712 USA}

\email{sdr@astro.as.utexas.edu}

\begin{abstract}
Previous studies of the interior structure of transiting exoplanets have
shown that the heavy element content of gas giants increases with host
star metallicity. Since metal-poor planets are less dense and have
larger radii than metal-rich planets of the same mass, one might expect
that metal-poor stars host a higher proportion of gas giants with large
radii than metal-rich stars. Here I present evidence for a negative
correlation at the $2.3 \sigma$ level between eclipse depth and stellar
metallicity in the {\it Kepler} gas giant candidates. Based on Kendall's
$\tau$ statistics, the probability that eclipse depth depends on star
metallicity is $0.981$. The correlation is consistent with planets
orbiting low-metallicity stars being, on average, larger in comparison
with their host stars than planets orbiting metal-rich stars.
Furthermore, since metal-rich stars have smaller radii than metal-poor
stars of the same mass and age, a uniform population of planets should
show a {\it rise} in median eclipse depth with [M/H]. The fact that I
find the opposite trend indicates that substantial changes in gas giant
interior structure must accompany increasing [M/H]. I
investigate whether the known scarcity of giant planets orbiting
low-mass stars could masquerade as an eclipse depth-metallicity
correlation, given the degeneracy between metallicity and temperature
for cool stars in the {\it Kepler} Input Catalog. While the eclise
depth-metallicity correlation is not yet on firm statistical footing and
will require spectroscopic [Fe/H] measurements for validation, it is an
intriguing window into how the interior structure of planets and even
the planet formation mechanism may be changing with Galactic chemical
evolution.
\end{abstract}


\section{Introduction}
\label{intro}

Since 1958, astronomers have known that Jupiter and Saturn are smaller
than they would be if they were made of pure hydrogen and helium
\citep{demarcus58}. Even though planetary radii depend on complex
equations of state, the simple principle that metal-rich planets are
more dense than their metal-poor counterparts of the same mass holds for
a wide range of planetary radii \citep[e.g.][]{fnett11}. Since the
heavy-element content of transiting gas giants increases with host star
metallicity \citep{guillot06, miller11}, one may make a simple
prediction: the gas giants orbiting metal-poor stars should have higher
radii, on average, than the gas giants orbiting metal-rich stars.  Just
as the planet detection rate depends on star metallicity
\citep{gonzalez98, santos04, fischer05}, the typical giant planet
interior structure also should change according to the metal content of
the host star.

Before the advent of the {\it Kepler} mission, there was no large,
uniform sample of gas giants available to search for a relationship
between stellar metallicity and gas giant size.  The surveys of
transiting gas giants were heterogeneous and mostly insensitive to
planets smaller than $\sim 0.7 R_{\rm Jup}$ (0.7 Jupiter radii). One
notable exception was GJ~436~b at $0.365 R_{\rm Jup}$, which orbits an
M2.5 star of radius $0.464 R_{\odot}$ \citep{gillon07b}. The {\it
Kepler} spacecraft, launched in March of 2009, is monitoring about
150,000 mostly Solar-type stars, primarily searching for transits of
Earthlike planets in the habitable zone \citep{batalha10}. Yet the {\it
Kepler} mission is also extraordinarily useful for studies of gas
giants, for the first time providing a sample of candidate giant planet
hosts observed under uniform conditions and with the sensitivity to
detect all of the Neptune-like objects of $\sim 0.3 R_{\rm Jup}$ and
below.

Here, then, is the opportunity to investigate the dependence of
planetary interior structure on host star metallicity. Of course, one
expects to see gas giants with a range of sizes orbiting all types
of Population 1 stars---after all, Jupiter and Neptune orbit the same
star but have radii that differ by almost a factor of three. However,
the {\it ensemble} of planet radii should show some dependence on
stellar metallicity if the heavy element contents of planets and their
parent stars are truly connected, as suggested by \citet{guillot06} and
\citet{miller11}. Likewise, since the {\it Kepler} mission is surveying
primarily similar, Sunlike stars, one also expects a negative
correlation between gas giant eclipse depth and star metallicity. In
this article I present the first evidence that such a trend
exists---though at this time the eclipse depth-metallicity trend is on
tenuous statistical footing and will require follow-up spectroscopic
observations for verification.

This article is organized as follows: In \S \ref{sample} I outline my gas
giant selection criteria. In \S \ref{statistics} I present statistical
support for a negative correlation between eclipse depth and star
metallicity. I discuss possible selection biases that could masquerade
as an eclipse depth-metallicity trend in \S \ref{biases}.  Finally, I
examine physical mechanisms besides the simple density effect that could
lead to an eclipse depth-metallicity correlation and speculate about the
implications for planet formation theory in \S \ref{physical}.

\section{Sample selection}
\label{sample}

In February 2011, the {\it Kepler} team released data for 1235 planet
candidates \citep{borucki11a}. While the planet/star radius ratios
$R_p/R_*$ were directly measured from eclipse depths, the values of
$R_*$ were based on $\log (g)$ measurements from the Kepler Input
Catalog \citep[KIC;][]{brown11}. The KIC stellar parameters were
computed using photometry in the Sloan {\it u, g, r, i, z} filters, the
intermediate-bandwidth $D51$ filter, and the 2MASS $JHK$ filters
\citep{skrutskie06}. \citet{brown11} quote a 0.4-dex error bar on $\log
(g)$, which translates into a +58\%/-37\% radius error for a planet
transiting a Sunlike star. Hotter, early- to mid-F stars have higher
$\log (g)$ errors, while subgiants tend to have systematically high
$\log (g)$ estimates. Furthermore, \citet{chaplin10} performed a
detailed asteroseismic follow-up study of three Sunlike stars and found
that the KIC $T_{\rm eff}$ estimates were systematically low.
\citet{pinsonneault12} also found that temperatures derived from KIC
$griz$ photometry were systematically lower than temperatures computed
by applying the infrared flux method \citep{casagrande10} to 2MASS
photometry \citep{skrutskie06}. The problems with $\log (g)$ for hot
stars may therefore apply to stars that have apparently Sunlike
temperatures according to the KIC. Quoted planet candidate radii from
the {\it Kepler} data release are subject to significant uncertainties.

However, the {\it Kepler} planet candidate radii can still be used to
identify a set of likely gaseous, giant planet candidates.  Here I
define giant planets as objects with bulk density $\rho \la
1.5$~g~cm$^{-3}$, for which $\sim 50$\% or more of the planetary volume
is due to the H/He atmosphere. According to the planetary interior
models of \citet{helled11}, this volume-based definition of giant planet
would include Uranus and Neptune, which are about 65\% H/He by volume.
\citet{rogers10} show that HAT-P-11b \citep{bakos10} and GJ~436b
\citep{gillon07, torres07} must also have gas mass and volume fractions
similar to Uranus and Neptune. HAT-P-11b, GJ~436b, Uranus and Neptune
have radii of $4.7 R_{\oplus}$, $4.2 R_{\oplus}$, $4.1 R_{\oplus}$, and
$4.0 R_{\oplus}$, respectively, suggesting that gas giants in general
have radii $R_p \ga 4 R_{\oplus}$. To be conservative, I follow
\citet{schlaufman11} and set the minimum planet candidate radius for
this study at $R_p = 5.0 R_{\oplus}$. For a $5 R_{\oplus}$ candidate
with a +58\% measurement error on planet size, the true size would be
$3.2 R_{\oplus}$, still within the size range where a gaseous atmosphere
is likely \citep{rogers10}.

The basic sample consists (1) of candidate giant planets with radius
$R_p \geq 5.0 R_{\oplus}$ that (2) also have metallicity measurements
from the KIC. Of the candidates that satisfy criteria (1) and (2), 28
have estimated sizes larger than the ``styrofoam planet'' Kepler-7b
\citep[$R_p = 18.1 R_{\oplus}$;][]{latham10}, the lowest-density
transiting planet yet known. Given that theoretical mass-radius
relations predict that Jupiter-mass ($M_{\rm Jup}$) planets, $20 M_{\rm
Jup}$ brown dwarfs and $100 M_{\rm Jup}$ stars have radii that vary by
less than 5\% \citep{chabrier09}, it is unfortunately very easy for
eclipsing low-mass companion stars to masquerade as giant planets. I
therefore carefully scrutinized the light curves of the largest planet
candidates for signs of (a) planet radius mismeasurement due to V-shaped
eclipses, (b) odd/even eclipse depth differences that indicate eclipsing
binaries (either in the target itself or a background object), and (c)
pulsations that suggest the $\delta$~Scuti-like variability of early- to
mid-F stars, for which the $\log (g)$ uncertainties are high.

Among the 23 candidates with metallicities from KIC and $R_p \geq 18.1
R_{\oplus}$, I find 7 objects for which the reported ratio $R_p / R_*$
was mis-fit by the {\it Kepler} data analysis pipeline. All such light
curves show V-shaped eclipses rather than flat-bottomed eclipses, which
drives the fitting algorithm toward artificially low impact parameters
and high values of $R_p / R_*$ (B.\ Cochran, private communication). Two
objects, KIC 5649956 and KIC 1432214, show the odd/even differences in
eclipse depth that indicate an eclipsing binary system. KIC 5649956
pulsates at the 3-4\% level, making the odd/even effect difficult to
identify. KIC 6470149 and KIC 7449844, which do not have odd/even
eclipse patterns, have been shown to host eclipsing M~dwarfs by
follow-up RV observations taken since the release of the {\it Kepler}
planet candidate list.  Meanwhile, follow-up spectra of KIC 5356593 have
raised its $\log g$ value from the 3.8 dex listed in the KIC to 4.5 dex,
correspondingly shrinking the computed planet radius from $35.5
R_{\oplus}$ to 13-14$R_{\oplus}$ (B.\ Cochran, private communication).

Furthering the list of suspicious objects, six other stars have no
obvious mismeasurement of eclipse depth, but their eclipses are
V-shaped, which is more consistent with a companion star than a planet.
One star, KIC 10616571, pulsates at the 3\% level on three frequencies,
while one or two small peaks occur at the minimum of each eclipse,
suggesting a hierarchical triple system. Finally, five stars that do not
have V-shaped eclipses or odd/even effects pulsate with similar
amplitude to the transit signal. One such pulsator, KIC 11818800, will
have its planet candidate radius revised from $36 R_{\oplus}$ to $9.77
R_{\oplus}$ in the upcoming re-release of the {\it Kepler} planet
candidate list (B.\ Cochran, private communication). The preponderance
of problems among the largest planet candidates suggest that one should
exercise caution when selecting a sample of inflated giant planets.

In a planet candidate list ordered according to $R_p$, I found that
pulsations, odd/even effects and mismeasured and/or V-shaped eclipses
accounted for fully 100\% of the objects with $R_p \ga 22 R_{\oplus}$.
Below $20 R_{\oplus}$, light curves that have no obvious problems begin
to dominate. The confirmed planet Kepler-12b has $R_p = 18.9 R_{\oplus}$
\citep{fortney11} and a similar density to Kepler-7b, validating the
idea that such inflated planets can exist. I therefore set an upper
limit of $20 R_{\oplus}$ to the planet candidates in this study. The
basic sample consists of 213 giant planet candidates with $5.0
R_{\oplus} < R_p < 20 R_{\oplus}$ whose host stars also have
metallicities from the Kepler Input Catalog (KIC). All planet
candidates used in this study are listed in Table 1.

While \citet{morton11} argue for a false positive rate of 5\%-10\% among
the {\it Kepler} planet candidates, \citet{borucki11a} are more cautious
and quote $> 80$\% reliability for rank 2 KOIs (Kepler Objects of
Interest) and $> 60$\% reliability for rank 3 and 4 KOIs. Despite the
scrutiny of planet candidates, there is therefore still the possibility
that the sample is contaminated by as much as 40\% eclipsing binaries,
background eclipsing binaries and hierarchical triples. I discuss the
possible consequences of such contamination in \S \ref{biases}.

\section{Statistical evidence for deep eclipses of low-metallicity stars}
\label{statistics}

With a sample of gas giant candidates in hand, I now present the first
statistical evidence that Galactic chemical evolution affects the
internal structure of giant planets as well as the frequency with which
they form. Figure \ref{scatterplot} shows $R_p / R_*$ as a function of
host star [M/H] for the stars in my sample. Planet candidates are
color-coded according to the irradiation they receive from their host
stars, which I discuss further in \S \ref{physical}. Although I used the
{\it Kepler} estimates of $R_p$ to select my sample of gas giant
candidates (see \S \ref{sample} for selection criteria), for the
purposes of statistical analysis I have cast the trend in terms of $R_p
/ R_*$ rather than $R_p$ because planet radii tend to change
substantially between inclusion in the {\it Kepler} planet candidate
list and eventual confirmation and publication. $R_p / R_*$, by
contrast, is a simple observable whose accuracy I have verified by
examining the light curves of the largest planet candidates. The
question I am asking is---are {\it Kepler} candidate giant planets
orbiting low-metallicity stars larger {\it in comparison to their host
stars} than candidate planets orbiting high-metallicity stars?

Figure \ref{scatterplot} shows eclipse depth as a function of [M/H] for
the candidate giant planet hosts in the selected sample. Scanning from
left to right, notice the knot of small planet candidates with $0.05 \la
R_p/R_* \la 0.13$ that appears at [M/H]~$ \ga -0.2$ without a
commensurate increase in large planet candidates with $R_p / R_* \ga
0.13$. To quantify the trend in decreasing eclipse depth with
metallicity, I have computed a running median \citep{hardle95} with a
21-point window (solid black) and a local polynomial regression
\citep{cleveland92} with span 0.75 and degree 1 (dash-dot black). Notice
the decreasing nature of both manifestations of the trend in the range
$-0.1 \la {\rm [M/H]} \la 0.3$.  While planets of all radii can appear
around stars of all metallicities, there is a hint that eclipses may be
deepest, on average, for low-metallicity stars. This possible trend is
as predicted---planets orbiting low-metallicity hosts should have the
lowest solid/gas ratios, making them the least dense and therefore
biggest of planets.

Given that the running median is noisy and the slope of the local
polynomial regression is small, is the trend statistically significant?
The Kendall rank coefficient for mapping [M/H] onto $R_p / R_*$ is $\tau
= -0.108$. The associated probability that the null hypothesis---that
$R_p / R_*$ is unrelated to star metallicity---is correct is $p =
0.019$, a significance level at which the R statistical suite recommends
adoption of the alternative hypothesis. For more information on
Kendall's $\tau$ coefficient, see \citet{abdi07}. Using the large-sample
Gaussian approximation to the variance $\nu$ of the $\tau$ sampling
distribution,
\begin{equation}
\nu = \frac{2 \left ( 2n + 5 \right )}{9n \left ( n - 1 \right
)},
\label{tauvariance}
\end{equation}
where $n = 213$ is the number of stars in the sample \citep{abdi07}, I
find that the standard deviation of the $\tau$ sampling distribution is
0.046.  The computed value of $\tau = -0.108$ differs from the $\tau =
0$ expected in the case of the null hypothesis by $-2.3 \sigma$. I
therefore consider the eclipse depth-metallicity trend interesting and
suggestive, but not conclusive---especially since so little is known
about the true distribution of giant planet eclipse depths that the
Gaussian approximation to the $\tau$ distribution may not be
appropriate. My main goal in writing this article is to motivate
spectroscopic follow-up of the {\it Kepler} giant planet candidates.
More precise metallicities and star radii will place our understanding
of how planet structure evolves with star metallicity on much firmer
statistical footing.

Besides the robust measures of center polotted in Figure
\ref{scatterplot} and the non-parametric Kendall's $\tau$ coefficient
discussed above, there is another way to visualize the eclipse
depth-metallicity trend.  In Figure \ref{histograms}, I have divided the
data into four broad metallicity bins and plotted the fraction of planet
candidates for which $R_p / R_* > 0.13$ in each bin. I chose $R_p / R_*
= 0.13$ as the cutoff because it is a roughly typical eclipse depth of
well-studied inflated planets such as HD~209458~b \citep[$R_p / R_* =
0.12$;][]{charbonneau00} and HAT-P-19~b \citep[$R_p / R_* =
0.14$;][]{hartman11}. What Figure \ref{histograms} also hints at, then,
is that stars in the lowest-metallicity bin tend to host a higher
fraction of planets that are large {\it in comparison to their host
stars} than the stars in the higher-metallicity bins. Notice, however,
that the Poisson error bars on the fraction of high eclipse depths in
the lowest-metallicity bin overlap with the error bars on two of the
other three bins. Furthermore, note that in binning the data and setting
a threshold value of $R_p / R_* = 0.13$, I have made the threshold and
the bins parameters that affect any statistical inferences made from
Figure \ref{histograms} (though slightly different cutoff values yield a
similar trend). Figure \ref{histograms} is intended more as a
visualization of the data than as a statistical tool.


There is one more potential pitfall left to examine, the accuracy of the
KIC [M/H] estimates. Already the planet size-metallicity trend is on
marginal footing---could [M/H] inaccuracies wipe it out entirely? Figure
\ref{accuracy} shows KIC [M/H] as a function of spectroscopically
determined [Fe/H]\footnote[1]{Host star metallicities for {\it Kepler}
discoveries were collected from the Extrasolar Planets Encyclopaedia,
http://exoplanet.eu.} for confirmed, published {\it Kepler} discoveries.
From Figure \ref{accuracy}, one can see that the KIC metallicities are
systematically lower than the spectroscopic values. The best-fit line
relating KIC [M/H] to spectroscopic [Fe/H] is
\begin{equation}
{\rm [M/H]_{KIC}} = 0.88 {\rm [Fe/H]_{spec}} - 0.14.
\label{metacc}
\end{equation}
This result agrees with the analysis of \citet{brown11}, whose
comparison of KIC metallicities with [Fe/H] derived from Keck/HIRES
spectra uncovered a systematic 0.17-dex underestimate in the KIC values.
The Spearman rank correlation coefficient relating ${\rm [Fe/H]_{spec}}$
and ${\rm [M/H]_{KIC}}$ is 0.76, which has a two-sided significance of
its deviation from zero of $2.7 \times 10^{-6}$.  In contrast,
\citet{brown11} found a Spearman rank correlation coefficient of only
0.42, with a significance of 0.02, for a subset of stars in the KIC
observed with Keck/HIRES. My analysis indicates that, despite the
differences in absolute scale between spectroscopic and KIC
metallicities, the two are nevertheless closely related. Note, however,
the two outliers, Kepler-10 and Kepler-19. Kepler-10 is a Solar-radius,
Solar-temperature star with G spectral type \citep{batalha11}, while
Kepler-19 is a slightly smaller star \citep[$R_* = 0.85
R_{\odot}$,][]{ballard11} with $T_{\rm eff} = 5541 \pm 60$~K and a
possibly later spectral type. Neither star falls into a known problem
category for the KIC---F stars, subgiants and stars with $T_{\rm eff} <
4200$~K---so there is no clear reason why their KIC metallicities are so
low. Published planet candidates used in the [M/H] accuracy
analysis are listed in Table \ref{mettable}.

Knowing that KIC metallicities do track spectroscopic metallicities,
albeit with some scatter, the relevant question for the purposes of this
study is whether or not one can place the {\it Kepler} candidate planet
hosts in the correct {\it order} on some metallicity scale.  Keeping in
mind the warnings about how binning introduces extra parameters into the
statistical analysis whose effects may not be well understood, I still
wish to know how often measurement error will cause a candidate planet
host to cross over into a neighboring bin in Figure \ref{histograms}.
The scatter of the KIC [M/H] values around the best-fit line in equation
\ref{metacc} is $\sigma_{\rm [M/H]} = 0.22$~dex, indicated by
dash-dotted lines in the left-hand panel of Figure \ref{accuracy}. (The
right-hand panel of Figure \ref{accuracy} shows KIC [M/H] as a function
of $T_{\rm eff}$, a correlation I will discuss further in \S
\ref{biases}.) I modeled the probability distribution of each star's
true metallicity using a Gaussian distribution with $\sigma = 0.22$~dex
centered on that star's KIC [M/H] value. These Gaussian distributions of
probable metallicity formed the basis of a Monte Carlo simulation, in
which I randomly sampled each star's [M/H] distribution 20,000 times,
divided the random samples into the same [M/H] bins as in Figure
\ref{histograms} and counted the fraction of planet candidates in each
bin with $R_p / R_* > 0.13$.

Figure \ref{montecarlo} shows the results of the Monte Carlo simulation.
Here the story from the error bars is slightly more promising. Candidate
planet hosts with $-0.85 \leq {\rm [M/H]} \leq -0.45$~dex show
significantly more deep eclipses for which $R_p / R_* > 0.13$ than stars
with ${\rm [Fe/H]} > -0.15$~dex. Once again, though, the error bars
barely miss overlapping, so that the statistical significance of the
trend is about $2 \sigma$---suggestive, but not definitive. Confirming
that gas giants orbiting low-metallicity stars really tend to have
larger radii than gas giants orbiting high-metallicity stars will
require extensive follow-up spectroscopy of the {\it Kepler} candidate
planet hosts. The California-Kepler survey is already underway and will
significantly improve stellar $\log (g)$, [Fe/H] and planet radius
estimates \citep{howard12}. With spectroscopic observations of a large
fraction of the giant planet candidates, it should even be possible to
cast the size-metallicity trend in terms of $R_p$ rather than $R_p /
R_*$.

\section{Statistical biases}
\label{biases}

In this section I discuss possible statistical biases that could
masquerade as a size-metallicity trend.  The first possible bias I wish
to examine is whether or not the {\it Kepler} mission is equally
sensitive to planets with minimum radius $5 R_{\oplus}$ in each
metallicity bin.  Given its ability to detect Earthlike planets
transiting Sunlike stars, {\it Kepler} has at least 25 times the
sensitivity required to detect planets with $R_p = 5 R_{\oplus}$
orbiting Solar-type stars, which produce eclipse depths of two parts per
thousand.  Planets with $R_p = 5 R_{\oplus}$ only become undetectable
when the star radius approaches $5 R_{\odot}$. Since the largest star
radius in the selected sample is $4.82 R_{\odot}$ and only five stars
have $R_* > 3 R_{\odot}$, there should be no systematic trend that
prevents detection of the smaller, $5 R_{\oplus}$ giant planets at low
metallicity.

Another possible bias arises from the interaction of two effects: the
scarcity of gas giants orbiting low-mass stars \citep[e.g.][]{bonfils11}
and the red colors of late K- and M-type stars. Among stars of a
constant $J-H$ color, those that are most metal-rich tend to have the
highest $g-r$ color because of the wealth of iron absorption lines in
the blue part of the spectrum \citep{schlaufman11}. Red optical color is
therefore an important metallicity indicator for stars in the KIC, but
its usefulness breaks down for late-type stars that are naturally red,
whatever their metallicity. \citet{brown11} show that the coolest stars,
those with $T_{\rm eff} \leq 4200$~K, tend to be artificially classified
with super-Solar metallicity. Above $T_{\rm eff} = 4200$~K, the
metallicity distribution is a well-behaved Gaussian that is not a
function of temperature. One might worry that the low proportion of
inflated giant planets orbiting the most metal-rich stars could be an
artifact of the known scarcity of gas giants orbiting low-mass stars
\citep{endl06, cumming08, bonfils11}. Perhaps low-mass stars whose
metallicities are artificially high account for much of the apparently
metal-rich part of the selected sample.  Fortunately, only four stars in
the sample have $T_{\rm eff} < 4200$, so the most problematic KIC
metallicities have largely been avoided.

Although the analysis of KIC atmospheric parameters by \citet{brown11}
does not show any obvious correlation between KIC [M/H] and $T_{\rm
eff}$ for stars with $T_{\rm eff} > 4200$ (see their Figure 11), I wish
to independently verify the lack of a temperature-metallicity
relationship. The right-hand panel of Figure \ref{accuracy} shows [M/H]
as a function of $T_{\rm eff}$ for stars in the selected sample. The
solid line indicates the mean [M/H] value and the dash-dot line gives
the best linear fit. The Spearman rank correlation coefficient is -0.19,
with a significance of 0.0049---three orders of magnitude higher than
the significance of the ${\rm [M/H]_{KIC}}$-${\rm [Fe/H]_{spec}}$
relationship (a high significance value indicates a weak correlation).
While the correlation analysis shows some evidence for a
metallicity-temperature relationship, KIC metallicity estimates appear
to be much more related to true star metallicities than temperatures. It
is likely that the eclipse depth-[M/H] correlation heralds a change in
planet structure with host star [M/H] rather than with host star mass
and temperature, but I am unwilling to completely rule out the
possibility that dearth of gas giants orbiting low-mass stars is
affecting the results.

Although the Kepler planet radii used to select the sample show no
correlation with $\log (g)$ (Figure \ref{loggfig}, left panel), a
possible degeneracy between [M/H] and $\log (g)$ in the KIC could affect
the results of this study. Red optical color is an indicator of both
high $\log (g)$ and high metallicity, and the only
intermediate-bandwidth filter used for KIC photometry---the $D51$
filter---is sensitive to both $\log (g)$ and [M/H]. Fortunately, the
correlation between $\log (g)$ and [M/H] is much weaker than the
correlation between $T_{\rm eff}$ and [M/H]. For [M/H] as a function of
$\log (g)$ (Figure \ref{loggfig}, right panel), I find a Spearman rank
correlation coefficient of -0.13, with a significance of 0.05.
Furthermore, the subgiant problem discussed earlier in this section
should apply to a small fraction of the stars in the sample: only seven
stars of 213, or 3\%, have $\log (g) < 0.7$.  As with temperature, I am
unwilling to definitively state that the weak degeneracy between KIC
$\log (g)$ and [M/H] plays no part in creating the eclipse
depth-metallicity correlation. However, I think it likely that declining
eclipse depths with metallicity are related to giant planet interior
structure rather than $\log (g)$-[M/H] degeneracies.

The final possible source of bias comes from the possible 5\%-40\%
contamination of the sample with eclipsing binaries, background
eclipsing binaries, and hierarchical triples. This bias arises because
stellar multiplicity rates depend on star mass. From $0.1 M_{\odot} - 40
M_{\odot}$, the binary fraction is an increasing function of star mass
\citep[e.g.][]{clark12}, with about 20\% of $0.1 M_{\odot}$ stars in
binaries and 40\% of $1 M_{\odot}$ stars in binaries. The increasing
binary fraction with $M_*$ may mean that the planet candidates orbiting
stars with the largest radii have the highest contamination rate. Since
the tendency in the KIC is for low-mass, low-temperature stars to have
artificially high metallicity values, it is possible for background
eclipsing binaries to produce an eclipse depth-metallicity trend.
However, the binary fraction is only a source of bias if [M/H] is
related to star temperature in the candidate giant planet sample. Given
that my analysis indicates that KIC metallicity is more related to
spectroscopic metallicity than any other stellar parameter, I do not
think the multiple-star contamination is the source of the possible
eclipse depth-metallicity trend---but, again, more follow-up is
necessary to rule out multiple stars as a source of bias.

\section{Physical explanations for the eclipse depth-metallicity
correlation}
\label{physical}

Assuming that the tentative correlation between planet candidate eclipse
depth and metallicity (1) is real, and (2) is not a manifestation of the
dependence of planet occurrence rate on stellar mass---both of which are
assumptions that require further investigation---what is the physical
reason for the correlation? First, it is important to point out that for
stars of a given mass and age, star radius {\it decreases} with
increasing metallicity \citep[e.g.][]{marigo08}---so the fall in median
$R_p/R_*$ with [M/H] means gas giant interior structure must change
substantially to counteract the fact that eclipse depth should tend to
{\it rise} with star metallicity for a uniform population of planets. To
the extent that the KIC [M/H] values track spectroscopic [Fe/H], the
eclipse depth-metallicity correlation reflects real changes in the
population of giant planets as a function of star metallicity.

\begin{figure}
\epsscale{0.7}
\plotone{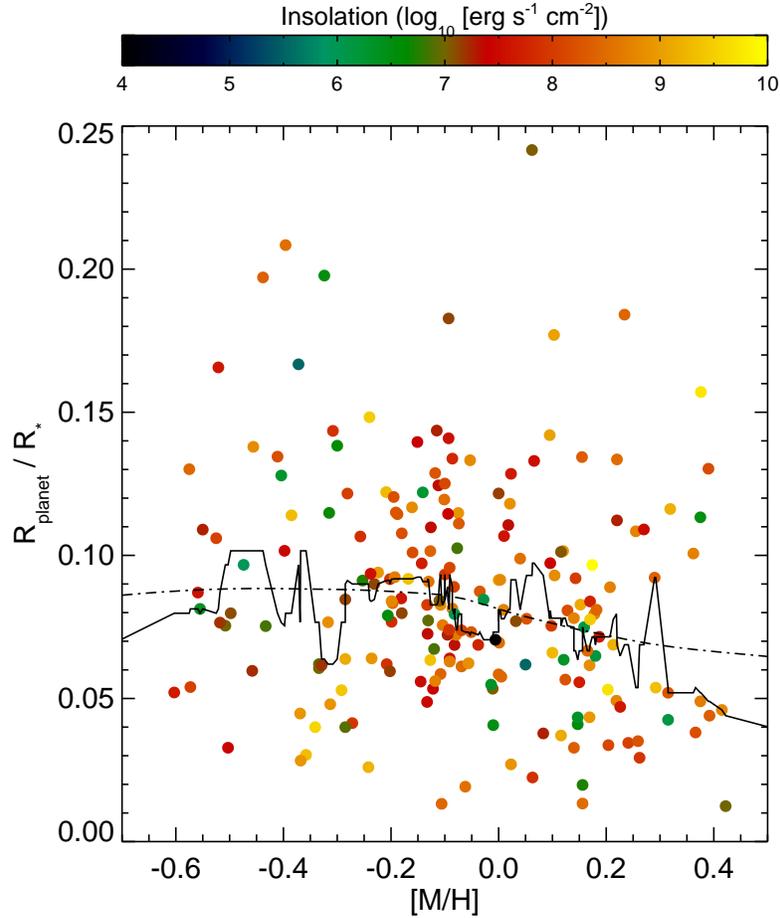}
\caption{An analysis of the depth of Kepler planet candidate eclipses
provides provides tentative support for the idea that planetary interior
structure evolves with cosmic time. Here I plot $R_p/R_*$ as a function
of host star [M/H]. The solid black line is a running median with a
21-point window, while the dash-dot line is a local polynomial
regression of span 0.75 and degree 1. Both robust measures of center
show a decreasing trend in eclipse depth with [M/H]. Points on each plot
are color-coded by level of stellar irradiation (``insolation'') at the
planet orbit. There is no indication that planets the orbiting
low-metallicity hosts are more heavily irradiated than other planets.}
\label{scatterplot}
\end{figure}

\begin{figure}
\epsscale{0.7}
\plotone{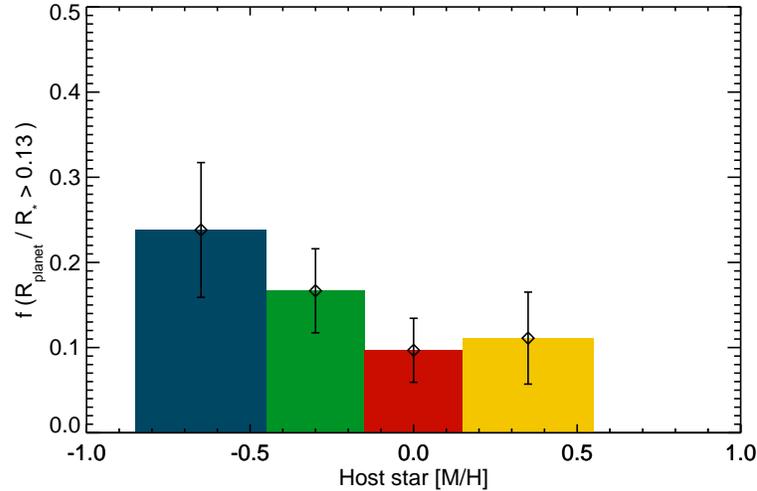}
\caption{Fraction of giant planet candidates with $R_p / R_*
> 0.13$. Here the planets orbiting the lowest-metallicity hosts appear
to be significantly bigger, relative to their parent stars, than the
planets orbiting Solar-metallicity hosts. }
\label{histograms}
\end{figure}

\begin{figure}
\epsscale{1.0}
\plottwo{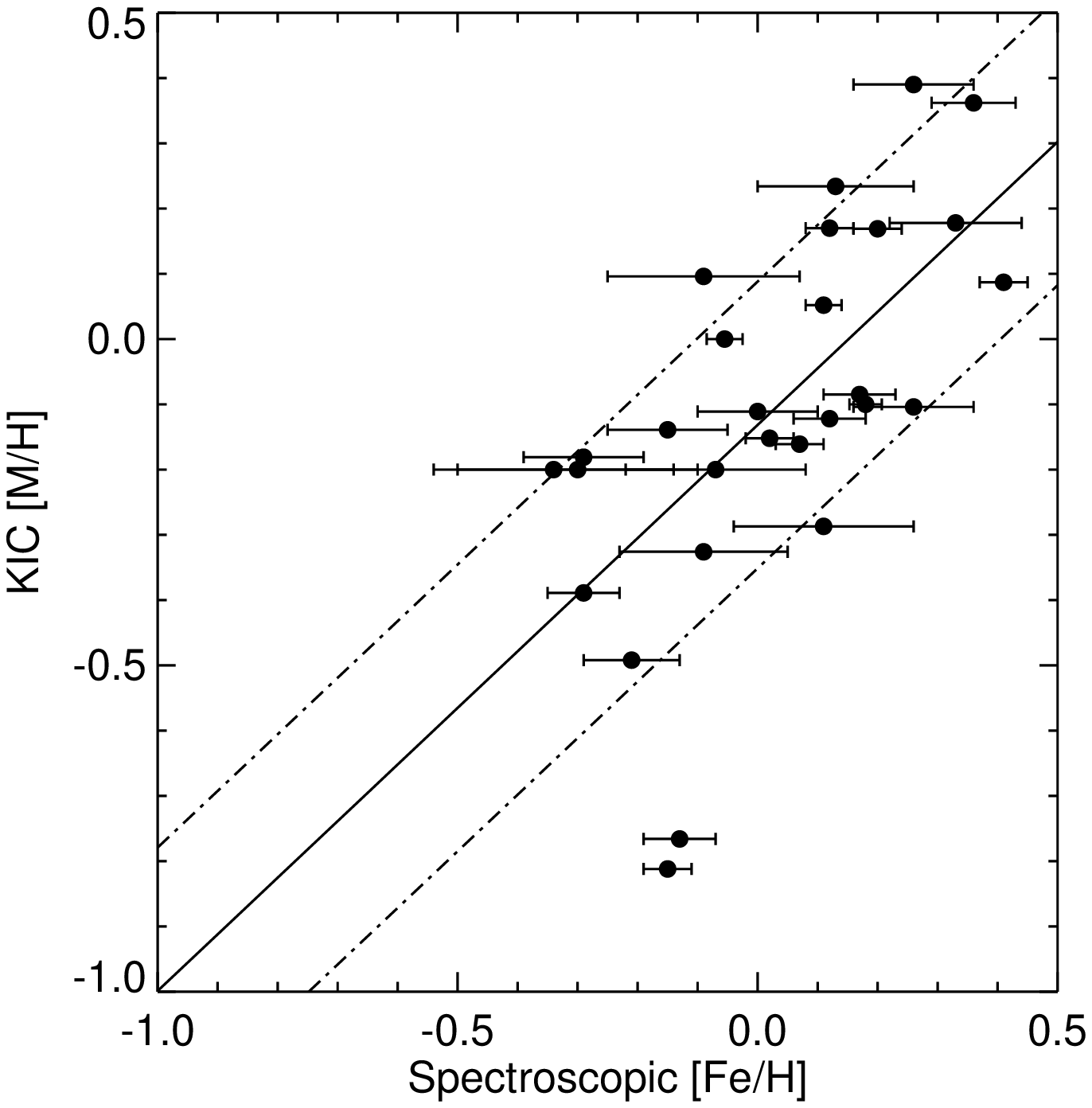}{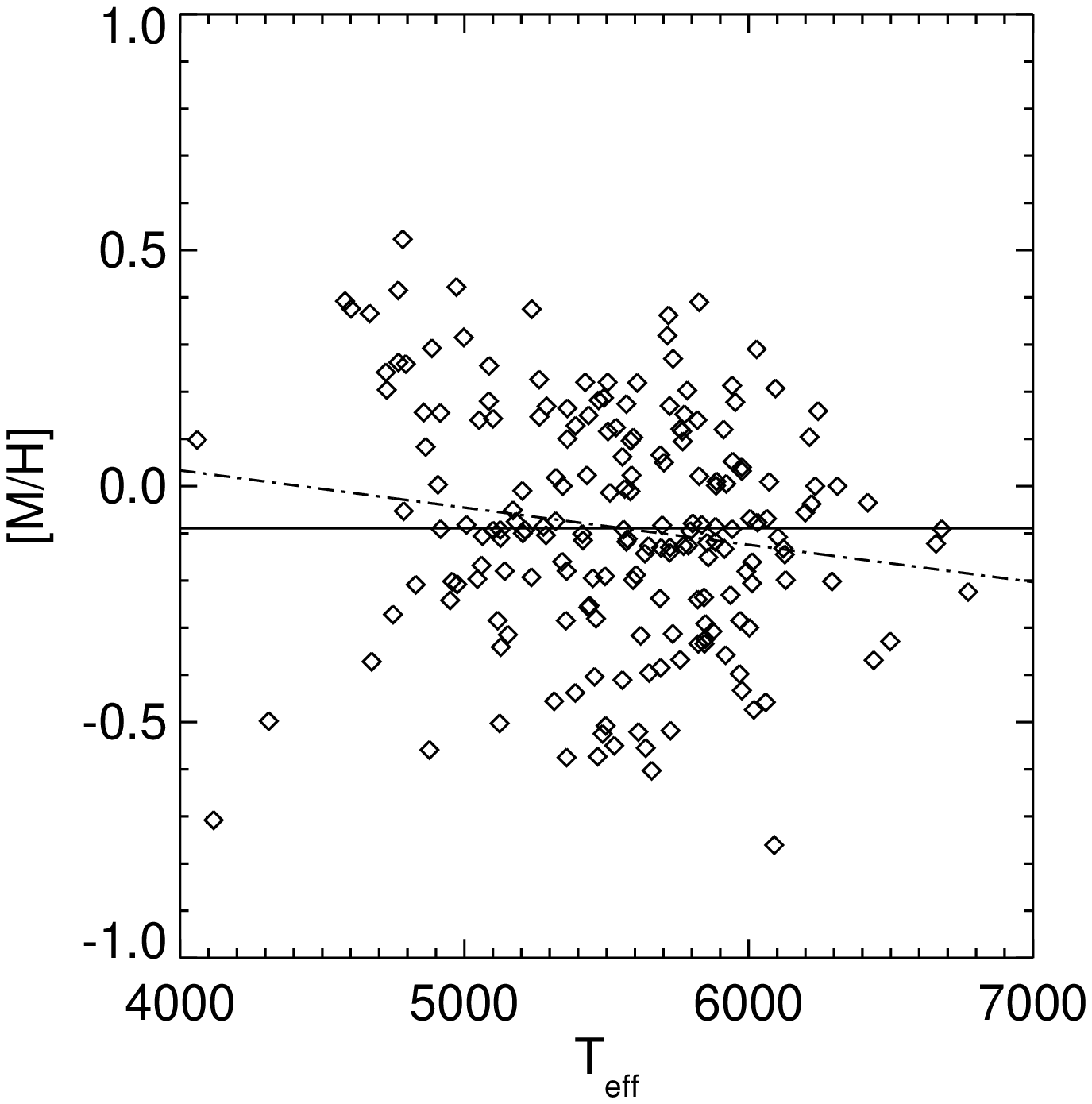}
\caption{{\bf Left:} A plot of [M/H] from the Kepler Input Catalog (KIC)
as a function of spectroscopially determined [M/H] for published Kepler
discoveries provides insight into the precision and accuracy of KIC
metallicities. The best-fit line (black) is $y = 0.88x - 0.14$,
indicating that KIC metallicities are systematically lower than their
spectroscopic counterparts. The {\it precision} of the KIC
metallicities, $\sigma_{\rm [M/H]} = 0.22$, is indicated by the
dash-dotted lines. The two outliers, Kepler-10 and Kepler-19, are both
Solar-type, single dwarf stars.  
{\bf Right:} Here I investigate whether a [M/H] is a function of
$T_{\rm eff}$ in the selected sample, a relationship that could
introduce statistical biases due to changes in both planet and
companion star occurrence rates as a function of stellar mass.
The solid line shows the mean [M/H] of the sample, while the
dash-dotted line shows the best-fit line representing [M/H] vs.
$T_{\rm eff}$. The Spearman rank correlation coefficient is
-0.19, indicating a weak dependence of metallicity on
temperature.}
\label{accuracy}
\end{figure}

\begin{figure}
\epsscale{1.0}
\plottwo{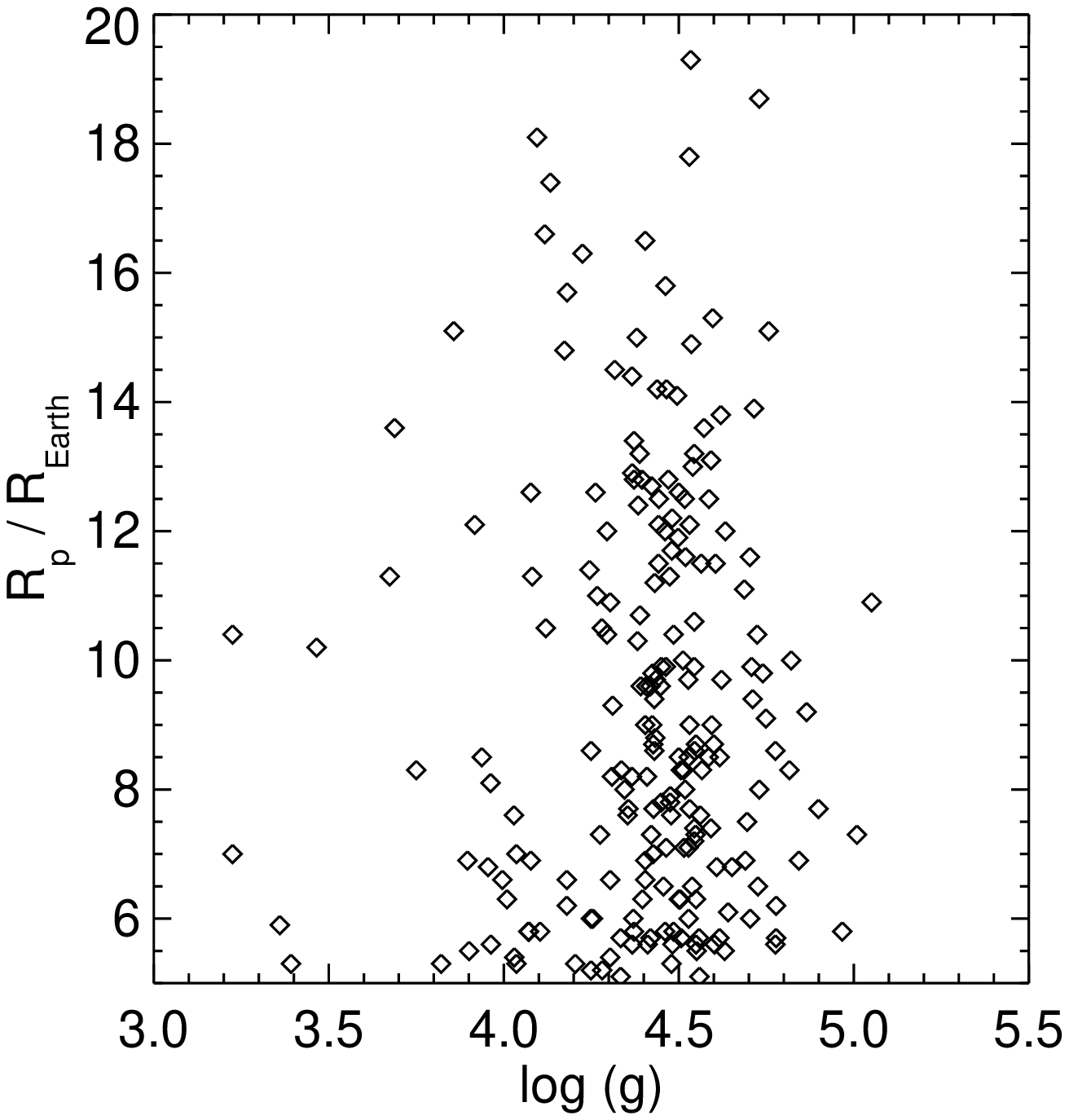}{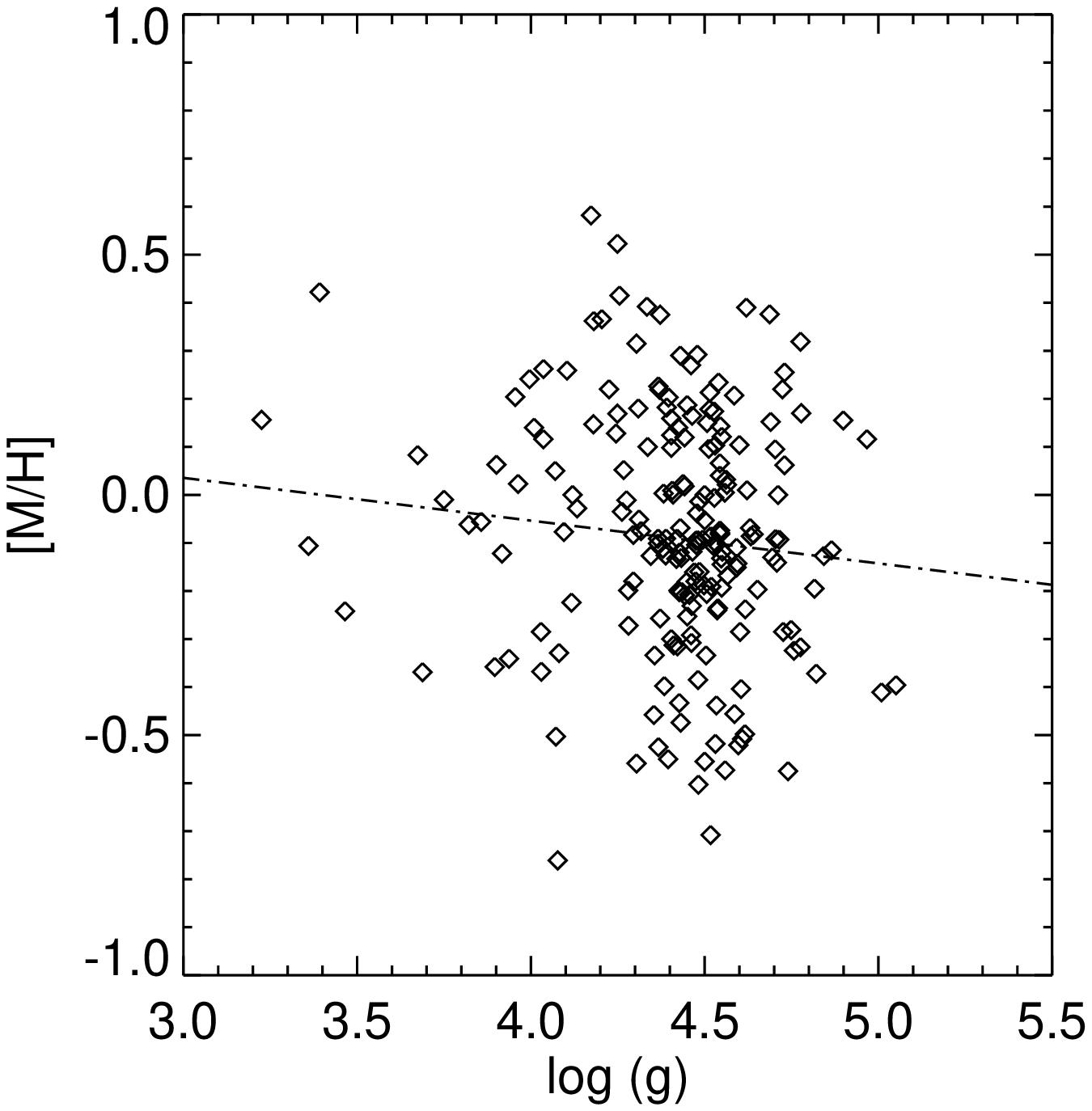}
\caption{{\bf Left:} The Spearman rank correlation coefficient
for planet radius as a function of $\log (g)$ is 0.04, with
significance 0.55, indicating no trend that could influence
sample selection. {\bf Right:} [M/H] and $\log (g)$ values in
the KIC are weakly correlated, with a Spearman rank coefficient
of -0.13. However, my analysis indicates that KIC [M/H] is more
strongly related to spectroscopic [Fe/H] than any other
parameter. The dash-dot line shows the best fit representing
[M/H] as a function of $\log (g)$.}
\label{loggfig}
\end{figure}

\begin{figure}
\epsscale{0.7}
\plotone{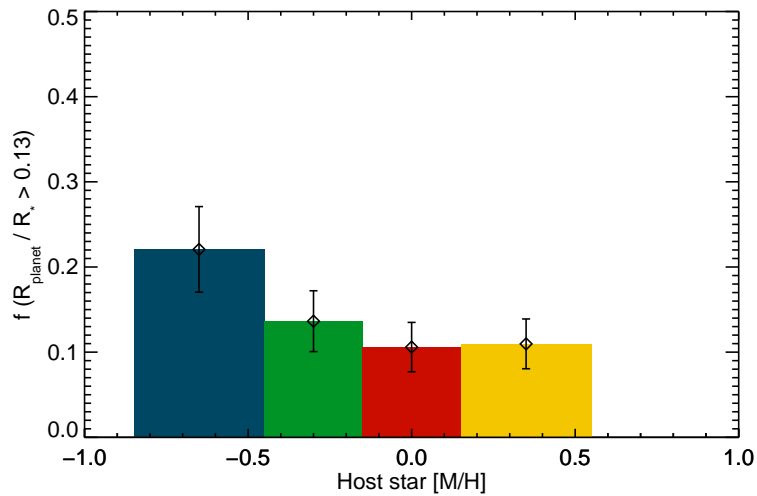}
\caption{Monte Carlo simulations that take into account the
errors in KIC metallicities suggest that there is a relationship between
gas giant size and host star [M/H]. Here I show the fraction of systems
with $R_p / R_* > 0.13$ as a function of host star [Fe/H] from $2 \times
10^4$ Monte Carlo simulations, assuming errors on KIC metallicities are
Gaussian and $\sigma_{\rm [M/H]} = 0.22$. The higher occurrence rate of
large planets around the lowest-metallicity hosts, compared with
Solar-metallicity and supersolar hosts, appears to be statistically
significant.}
\label{montecarlo}
\end{figure}

One possibility is that irradiation from the central star depends on
metallicity. Irradiation provides the energy required to inflate hot
Jupiters either through Ohmic dissipation \citep[which depends on the
star's ability to ionize the upper atmosphere;][]{batygin11} or simply
by forcing a shallow atmospheric temperature gradient that slows
planetary contraction \citep[e.g.][]{guillot02, baraffe03, burrows03,
fnett11}. At a given mass, a star of low metallicity will be slightly
bluer and hotter than its counterpart at high metallicity, so metal-poor
stars might tend to host larger-radius planets if the orbital
distribution does not depend on metallicity. The points in Figure
\ref{scatterplot} are color-coded according to average stellar
insolation.  Although there is a visible tendency for least-irradiated
planets to have the smallest eclipse depths (note the cluster of green
points toward the bottom of the plot), there is no obvious tendency for
the planets orbiting the lowest-metallicity hosts to be the most
irradiated. I therefore tentatively rule out differences in irradiation
levels as the reason for the eclipse depth-metallicity correlation.

Other planetary inflation mechanisms that have been proposed include
tidal heating \citep{jackson08}, thermal tides \citep{arras10} and
double-diffusive convection \citep{chabrier07}. Of these alternative
inflation mechanisms, double-diffusive convection---in which composition
gradients create multiple, semi-detached convective layers between which
energy transport is inefficient---is the most likely to be related to
star metallicity. In their models of the interior structures of
uninflated hot Jupiters, \citet{miller11} found a positive correlation
between stellar metallicity and heavy-element mass in planetary
envelopes (see their Figure 2). A metal-rich planetary envelope would
likely have a steeper composition gradient than a metal-poor envelope.
It is therefore possible that double-diffusive convection
provides the required connection between giant planet structure
and stellar metallicity.

The most straightforward explanation for the eclipse depth-metallicity
trend, however, is that metal-rich planets of a given mass are denser
than their metal-poor counterparts, leading to smaller radii
\citep[e.g.][]{fnett11}. Transit searches have for some time provided
hints that heavily metal-enriched, dense planets tend to orbit
metal-rich stars such as HD~149026 \citep{sato05} and HAT-P-2
\citep{bakos07, leconte09}, and the connection between heavy-element
mass and host star metallicity---the crucial ingredient behind any
eclipse depth-[M/H] correlation---has also been statistically
established \citep{guillot06, miller11}.  Given that Jupiter, Saturn,
Uranus and Neptune all orbit the same star but have radii ranging from
$4.0 R_{\oplus}$ to $11.2 R_{\oplus}$, one expects to see gas giants
with a range of radii orbiting stars of all metallicities yet probed
(whether or not giant planets should form around Population II stars is
an interesting question). However, if Jupiter, Saturn, Uranus and
Neptune were all twice as metal-rich as they are, the average radius of
the entire {\it ensemble} of Solar System planets would decline. I favor
a simple density effect as the reason behind the increased eclipse
depths of planet candidates orbiting low-metallicity stars.

I now make a speculative proposition, which is that low-metallicity
stars may host a higher proportion of planets that formed by
gravitational instability than high-metallicity stars. While dust is the
necessary raw ingredient for the protoplanetary cores that nucleate the
growth of massive gaseous atmospheres in the core accretion scenario,
dust is the enemy of planet formation by gravitational instability
because it increases disk opacity, reducing the disk's ability to cool
\citep{cai06}. Though planets that form by core accretion tend to be
metal-rich even relative to their host stars and by necessity have solid
cores \citep{miller11}, planets that form by gravitational instability
most likely have the stellar composition and tend to be less dense
\citep[see, however,][]{helled10}. The idea that the giant planets
orbiting low-metallicity stars predominantly formed by gravitational
instability is interesting and warrants further thought.  The theory
that high-metallicity stars and low-metallicity stars form their gas
giants by different mechanisms is consistent with the results of
\citet{santos04} and \citet{udry07}, who found that the frequency of
giant planets as a function of [Fe/H] is flat at subsolar metallicities.
If the dominant planet formation mechanism were not changing with
metallicity, one would expect a monotonically increasing planet
detection rate with [Fe/H] for the entire metallicity range surveyed.

While the eclipse depth-metallicity trend is not yet on firm statistical
footing, it is an intriguing possibility that may provide insight into
the Galactic history of planet formation. This work provides the first
piece of evidence that giant planet {\it structure}, not just detection
rate, depends on stellar metallicity. The planet populations of
metal-rich stars and metal-poor stars may be very different.  It is even
possible that differences in planetary structure with host star
metallicity could indicate not only changes in the raw ingredients
available for planet formation as the Galaxy becomes chemically
enriched, but changes in the planet formation mechanism
itself---planetary evolution across cosmic time.

Funding for this work was provided by the National Science Foundation
through the Faculty Early Career Development (CAREER) program, award
AST-1055910, to S. Dodson-Robinson. I thank Bill Cochran for input his
extensive help with weeding out suspicious giant planet candidates from
the {\it Kepler} database, Eric Feigelson for advice on nonparametric,
robust statistical methods, and the anonymous referee for a thorough
critique of the manuscript. This research has made use of the NASA
Exoplanet Archive, the Exoplanet Encyclopedia and the SIMBAD database,
operated at CDS, Strasbourg, France.

\clearpage
\begin{center}
\begin{longtable*}{rrrrr}
\caption{Giant planet candidates used in analysis
\label{startable} } \\

\hline \hline
KOI number\footnotemark[1] & KIC number &
$R_p/R_*$ & $R_p/R_*$ error & KIC [M/H] \\
\hline
\endhead

\hline
\hline
\endfoot

K00005.01 & 8554498 & 0.037 & 0.0002 &  0.12 \\ 
K00010.01 & 6922244 & 0.091 & 0.0007 &  0.00 \\ 
K00012.01 & 5812701 & 0.087 & 0.0001 & -0.04 \\ 
K00020.01 & 11804465 & 0.117 & 0.0000 & -0.16 \\ 
K00022.01 & 9631995 & 0.092 & 0.0001 &  0.29 \\ 
K00063.01 & 11554435 & 0.057 & 0.0010 &  0.12 \\ 
K00064.01 & 7051180 & 0.040 & 0.0003 & -0.34 \\ 
K00082.01 & 10187017 & 0.034 & 0.0025 &  0.20 \\ 
K00089.02 & 8056665 & 0.022 & 0.0006 &  0.06 \\ 
K00094.01 & 6462863 & 0.070 & 0.0024 & -0.76 \\ 
K00094.03 & 6462863 & 0.038 & 0.0016 & -0.76 \\ 
K00097.01 & 5780885 & 0.078 & 0.0001 &  0.05 \\ 
K00098.01 & 10264660 & 0.053 & 0.0094 & -0.12 \\ 
K00100.01 & 4055765 & 0.045 & 0.0005 & -0.37 \\ 
K00102.01 & 8456679 & 0.030 & 0.0024 & -0.36 \\ 
K00105.01 & 8711794 & 0.039 & 0.0006 & -1.08 \\ 
K00113.01 & 2306756 & 0.066 & 0.0210 &  0.10 \\ 
K00127.01 & 8359498 & 0.097 & 0.0001 &  0.17 \\ 
K00128.01 & 11359879 & 0.101 & 0.0005 &  0.36 \\ 
K00131.01 & 7778437 & 0.075 & 0.0001 &  0.16 \\ 
K00135.01 & 9818381 & 0.080 & 0.0074 &  0.18 \\ 
K00137.01 & 8644288 & 0.043 & 0.0002 &  0.17 \\ 
K00137.02 & 8644288 & 0.062 & 0.0013 &  0.17 \\ 
K00138.01 & 8506766 & 0.094 & 0.0006 & -0.22 \\ 
K00139.01 & 8559644 & 0.058 & 0.0022 &  0.01 \\ 
K00144.01 & 4180280 & 0.035 & 0.0047 &  0.24 \\ 
K00161.01 & 5084942 & 0.029 & 0.0068 &  0.26 \\ 
K00183.01 & 9651668 & 0.122 & 0.0190 & -0.14 \\ 
K00186.01 & 12019440 & 0.118 & 0.0001 &  0.02 \\ 
K00187.01 & 7023960 & 0.142 & 0.0130 &  0.10 \\ 
K00188.01 & 5357901 & 0.108 & 0.0001 &  0.26 \\ 
K00189.01 & 11391018 & 0.133 & 0.0008 & -0.05 \\ 
K00190.01 & 5771719 & 0.112 & 0.0006 &  0.22 \\ 
K00191.01 & 5972334 & 0.115 & 0.0011 & -0.19 \\ 
K00192.01 & 7950644 & 0.090 & 0.0210 & -0.23 \\ 
K00193.01 & 10799735 & 0.129 & 0.0002 & -0.12 \\ 
K00194.01 & 10904857 & 0.134 & 0.0037 & -0.09 \\ 
K00195.01 & 11502867 & 0.114 & 0.0019 & -0.19 \\ 
K00196.01 & 9410930 & 0.097 & 0.0009 &  0.10 \\ 
K00197.01 & 2987027 & 0.091 & 0.0001 &  0.00 \\ 
K00199.01 & 10019708 & 0.093 & 0.0170 &  0.10 \\ 
K00200.01 & 6046540 & 0.083 & 0.0001 &  0.15 \\ 
K00201.01 & 6849046 & 0.072 & 0.0001 &  0.19 \\ 
K00202.01 & 7877496 & 0.102 & 0.0005 &  0.12 \\ 
K00203.01 & 10619192 & 0.130 & 0.0002 &  0.39 \\ 
K00204.01 & 9305831 & 0.076 & 0.0001 & -0.10 \\ 
K00205.01 & 7046804 & 0.092 & 0.0020 & -0.17 \\ 
K00206.01 & 5728139 & 0.063 & 0.0001 & -0.13 \\ 
K00208.01 & 3762468 & 0.089 & 0.0007 &  0.21 \\ 
K00209.01 & 10723750 & 0.069 & 0.0002 & -0.04 \\ 
K00211.01 & 10656508 & 0.081 & 0.0960 &  0.01 \\ 
K00212.01 & 6300348 & 0.064 & 0.0140 & -0.24 \\ 
K00214.01 & 11046458 & 0.111 & 0.0035 &  0.02 \\ 
K00216.01 & 6152974 & 0.065 & 0.0004 &  0.18 \\ 
K00217.01 & 9595827 & 0.134 & 0.0001 &  0.22 \\ 
K00229.01 & 3847907 & 0.049 & 0.0002 &  0.22 \\ 
K00242.01 & 3642741 & 0.056 & 0.0002 &  0.15 \\ 
K00254.01 & 5794240 & 0.184 & 0.0012 &  0.23 \\ 
K00256.01 & 11548140 & 0.123 & 0.0023 &  0.58 \\ 
K00261.01 & 5383248 & 0.027 & 0.0067 &  0.02 \\ 
K00345.01 & 11074541 & 0.035 & 0.0063 &  0.26 \\ 
K00348.01 & 11194032 & 0.038 & 0.0003 &  0.37 \\ 
K00351.01 & 11442793 & 0.083 & 0.0064 & -0.11 \\ 
K00351.02 & 11442793 & 0.059 & 0.0006 & -0.11 \\ 
K00353.01 & 11566064 & 0.064 & 0.0012 & -0.09 \\ 
K00356.01 & 11624249 & 0.033 & 0.0071 & -0.50 \\ 
K00367.01 & 4815520 & 0.044 & 0.0085 & -1.53 \\ 
K00368.01 & 6603043 & 0.085 & 0.0001 & -0.03 \\ 
K00372.01 & 6471021 & 0.081 & 0.0093 & -0.56 \\ 
K00375.01 & 12356617 & 0.077 & \nodata & -0.13 \\ 
K00377.01 & 3323887 & 0.078 & 0.0030 &  0.17 \\ 
K00377.02 & 3323887 & 0.084 & 0.0017 &  0.17 \\ 
K00398.01 & 9946525 & 0.092 & 0.0036 &  0.14 \\ 
K00401.01 & 3217264 & 0.041 & 0.0003 &  0.15 \\ 
K00401.02 & 3217264 & 0.043 & 0.0034 &  0.15 \\ 
K00410.01 & 5449777 & 0.102 & 0.0050 & -0.40 \\ 
K00412.01 & 5683743 & 0.053 & 0.0002 & -0.01 \\ 
K00415.01 & 6289650 & 0.062 & 0.0630 & -0.33 \\ 
K00417.01 & 6879865 & 0.097 & 0.0017 & -0.14 \\ 
K00418.01 & 7975727 & 0.115 & 0.0004 & -0.32 \\ 
K00419.01 & 8219673 & 0.091 & 0.0004 & -0.13 \\ 
K00421.01 & 9115800 & 0.115 & 0.0002 & -0.07 \\ 
K00422.01 & 9214713 & 0.138 & \nodata & -0.30 \\ 
K00423.01 & 9478990 & 0.085 & 0.0003 & -0.18 \\ 
K00425.01 & 9967884 & 0.133 & 0.0140 &  0.07 \\ 
K00428.01 & 10418224 & 0.056 & 0.0001 & -0.14 \\ 
K00433.01 & 10937029 & 0.049 & 0.0180 &  0.38 \\ 
K00433.02 & 10937029 & 0.113 & 0.0013 &  0.38 \\ 
K00458.01 & 7504328 & 0.077 & 0.0062 & -0.20 \\ 
K00464.01 & 8890783 & 0.067 & 0.0003 &  0.17 \\ 
K00469.01 & 9703198 & 0.061 & 0.0029 & -0.07 \\ 
K00523.01 & 8806123 & 0.063 & 0.0015 & -0.09 \\ 
K00552.01 & 5122112 & 0.097 & 0.0013 & -0.47 \\ 
K00554.01 & 5443837 & 0.069 & 0.0034 & -0.08 \\ 
K00607.01 & 5441980 & 0.075 & 0.0009 & -0.51 \\ 
K00609.01 & 5608566 & 0.089 & 0.0110 & -0.08 \\ 
K00611.01 & 6309763 & 0.073 & 0.0004 & -0.13 \\ 
K00617.01 & 9846086 & 0.177 & 0.0210 &  0.10 \\ 
K00620.01 & 11773022 & 0.072 & 0.0009 & -0.08 \\ 
K00622.01 & 12417486 & 0.073 & 0.0025 & -0.05 \\ 
K00625.01 & 4449034 & 0.062 & 0.0097 & -0.06 \\ 
K00633.01 & 4841374 & 0.028 & 0.0023 & -0.37 \\ 
K00674.01 & 7277317 & 0.038 & 0.0042 &  0.08 \\ 
K00680.01 & 7529266 & 0.060 & 0.0001 & -0.46 \\ 
K00684.01 & 7730747 & 0.041 & 0.0042 & -0.01 \\ 
K00686.01 & 7906882 & 0.108 & 0.0032 & -0.18 \\ 
K00716.01 & 9846348 & 0.061 & 0.0027 & -0.33 \\ 
K00725.01 & 10068383 & 0.083 & 0.0024 & -0.20 \\ 
K00728.01 & 10221013 & 0.099 & 0.0010 &  0.04 \\ 
K00737.01 & 10345478 & 0.064 & 0.0020 & -0.28 \\ 
K00741.01 & 10418797 & 0.242 & 0.0062 &  0.06 \\ 
K00743.01 & 10464078 & 0.087 & 0.0210 & -0.56 \\ 
K00745.01 & 10485250 & 0.092 & 0.0006 & -0.20 \\ 
K00753.01 & 10811496 & 0.102 & 0.0072 & -0.13 \\ 
K00760.01 & 11138155 & 0.107 & 0.0008 &  0.01 \\ 
K00763.01 & 11242721 & 0.110 & 0.0009 & -0.13 \\ 
K00764.01 & 11304958 & 0.047 & 0.0004 &  0.23 \\ 
K00767.01 & 11414511 & 0.128 & 0.0008 &  0.02 \\ 
K00771.01 & 11465813 & 0.124 & \nodata & -0.11 \\ 
K00772.01 & 11493732 & 0.070 & 0.0044 &  0.00 \\ 
K00774.01 & 11656840 & 0.143 & 0.0004 & -0.31 \\ 
K00779.01 & 11909839 & 0.109 & 0.0210 & -0.55 \\ 
K00782.01 & 11960862 & 0.048 & 0.0930 & -0.31 \\ 
K00791.01 & 12644822 & 0.071 & 0.0003 & -0.01 \\ 
K00797.01 & 3115833 & 0.077 & 0.0007 & -0.52 \\ 
K00801.01 & 3351888 & 0.081 & 0.0130 &  0.18 \\ 
K00802.01 & 3453214 & 0.135 & 0.0049 & -0.41 \\ 
K00805.01 & 3734868 & 0.119 & 0.0095 & -0.10 \\ 
K00806.01 & 3832474 & 0.093 & 0.0073 & -0.10 \\ 
K00806.02 & 3832474 & 0.125 & 0.0005 & -0.10 \\ 
K00809.01 & 3935914 & 0.114 & 0.0650 & -0.39 \\ 
K00813.01 & 4275191 & 0.085 & 0.0003 & -0.28 \\ 
K00815.01 & 4544670 & 0.101 & 0.0130 & -0.16 \\ 
K00822.01 & 5077629 & 0.128 & 0.0015 & -0.40 \\ 
K00823.01 & 5115978 & 0.075 & 0.0015 & -0.43 \\ 
K00824.01 & 5164255 & 0.122 & 0.0013 & -0.21 \\ 
K00830.01 & 5358624 & 0.134 & 0.0001 &  0.15 \\ 
K00838.01 & 5534814 & 0.072 & 0.0010 & -0.10 \\ 
K00840.01 & 5651104 & 0.096 & 0.0031 & -0.09 \\ 
K00843.01 & 5881688 & 0.053 & 0.0049 &  0.20 \\ 
K00846.01 & 6061119 & 0.166 & 0.0008 & -0.52 \\ 
K00847.01 & 6191521 & 0.054 & 0.0330 & -0.57 \\ 
K00850.01 & 6291653 & 0.092 & 0.0042 & -0.19 \\ 
K00851.01 & 6392727 & 0.056 & 0.0007 & -0.12 \\ 
K00855.01 & 6522242 & 0.138 & 0.0002 & -0.46 \\ 
K00856.01 & 6526710 & 0.140 & 0.0025 & -0.15 \\ 
K00858.01 & 6599919 & 0.091 & 0.0027 & -0.25 \\ 
K00865.01 & 6862328 & 0.074 & 0.0360 & -0.09 \\ 
K00868.01 & 6867155 & 0.161 & 0.0025 & -0.71 \\ 
K00871.01 & 7031517 & 0.208 & 0.0044 & -0.40 \\ 
K00872.01 & 7109675 & 0.084 & 0.0027 & -0.11 \\ 
K00876.01 & 7270230 & 0.144 & 0.0066 & -0.12 \\ 
K00878.01 & 7303253 & 0.041 & 0.0031 & -0.27 \\ 
K00880.02 & 7366258 & 0.055 & 0.0004 & -0.01 \\ 
K00882.01 & 7377033 & 0.151 & 0.0015 & -1.24 \\ 
K00883.01 & 7380537 & 0.167 & 0.0005 & -0.37 \\ 
K00889.01 & 757450 & 0.114 & 0.0003 & -0.09 \\ 
K00890.01 & 7585481 & 0.077 & 0.0003 &  0.03 \\ 
K00895.01 & 7767559 & 0.107 & 0.0011 & -0.26 \\ 
K00897.01 & 7849854 & 0.109 & 0.0001 &  0.27 \\ 
K00902.01 & 8018547 & 0.080 & 0.0007 & -0.50 \\ 
K00903.01 & 8039892 & 0.077 & 0.0002 & -0.32 \\ 
K00908.01 & 8255887 & 0.081 & 0.0002 &  0.13 \\ 
K00913.01 & 8544996 & 0.122 & 0.0049 & -0.28 \\ 
K00918.01 & 8672910 & 0.111 & 0.0003 & -0.07 \\ 
K00929.01 & 9141746 & 0.078 & 0.0002 &  0.14 \\ 
K00931.01 & 9166862 & 0.116 & 0.0075 &  0.32 \\ 
K00941.01 & 9480189 & 0.043 & 0.0005 &  0.32 \\ 
K00941.03 & 9480189 & 0.052 & 0.0120 &  0.32 \\ 
K00951.01 & 9775938 & 0.046 & 0.0120 &  0.41 \\ 
K00956.01 & 9875711 & 0.044 & 0.0170 &  0.39 \\ 
K00960.01 & 8176650 & 0.183 & 0.0005 & -0.09 \\ 
K00972.01 & 11013201 & 0.019 & 0.0025 & -0.06 \\ 
K00981.01 & 8607720 & 0.013 & 0.0012 & -0.11 \\ 
K00988.01 & 2302548 & 0.033 & 0.0021 &  0.14 \\ 
K01003.01 & 2438502 & 0.141 & 0.0007 & -0.09 \\ 
K01005.01 & 5780460 & 0.062 & 0.0430 & -0.21 \\ 
K01089.01 & 3247268 & 0.083 & 0.0003 & -0.13 \\ 
K01089.02 & 3247268 & 0.049 & 0.0060 & -0.13 \\ 
K01159.01 & 10354039 & 0.054 & 0.0039 &  0.29 \\ 
K01176.01 & 3749365 & 0.157 & 0.0003 &  0.38 \\ 
K01177.01 & 3547091 & 0.130 & 0.0030 & -0.57 \\ 
K01193.01 & 3942446 & 0.106 & 0.0380 & -0.53 \\ 
K01208.01 & 3962440 & 0.060 & \nodata & -0.20 \\ 
K01221.02 & 3640905 & 0.012 & 0.0059 &  0.42 \\ 
K01227.01 & 6629332 & 0.120 & 0.0013 & -0.20 \\ 
K01241.01 & 6448890 & 0.020 & 0.0039 &  0.16 \\ 
K01241.02 & 6448890 & 0.013 & 0.0051 &  0.16 \\ 
K01242.01 & 6607447 & 0.058 & 0.0006 &  0.00 \\ 
K01257.01 & 8751933 & 0.080 & 0.0005 & -0.18 \\ 
K01261.01 & 8678594 & 0.064 & 0.0005 &  0.12 \\ 
K01268.01 & 8813698 & 0.074 & \nodata & -0.07 \\ 
K01285.01 & 10599397 & 0.081 & 0.0029 & -0.09 \\ 
K01288.01 & 10790387 & 0.084 & 0.0005 & -0.20 \\ 
K01299.01 & 10864656 & 0.026 & 0.0002 & -0.24 \\ 
K01335.01 & 4155328 & 0.040 & 0.0210 & -0.28 \\ 
K01353.01 & 7303287 & 0.102 & 0.0011 & -0.08 \\ 
K01385.01 & 9278553 & 0.198 & 0.0003 & -0.32 \\ 
K01391.01 & 8958035 & 0.079 & 0.0016 & -0.21 \\ 
K01419.01 & 11125936 & 0.053 & 0.0035 & -0.29 \\ 
K01426.02 & 11122894 & 0.067 & 0.0028 & -0.12 \\ 
K01459.01 & 9761199 & 0.075 & 0.0010 &  0.10 \\ 
K01474.01 & 12365184 & 0.062 & 0.0005 & -0.33 \\ 
K01477.01 & 7811397 & 0.122 & \nodata &  0.00 \\ 
K01486.01 & 7898352 & 0.094 & \nodata & -0.24 \\ 
K01540.01 & 5649956 & 0.197 & 0.0011 & -0.44 \\ 
K01543.01 & 5270698 & 0.148 & 0.0003 & -0.24 \\ 
K01546.01 & 5475431 & 0.101 & 0.0008 &  0.12 \\ 
K01553.01 & 7951018 & 0.069 & 0.0006 &  0.21 \\ 
K01557.01 & 5371776 & 0.039 & 0.0350 &  0.52 \\ 
K01561.01 & 4940438 & 0.052 & 0.0016 & -0.60 \\ 
K01574.01 & 10028792 & 0.062 & 0.0003 &  0.05 \\ 
K01587.01 & 9932970 & 0.080 & 0.0067 & -0.08 \\ 
\footnotetext[1]{Kepler Object of Interest number}
\end{longtable*}
\end{center}

\clearpage

\begin{deluxetable*}{lrrc}
\tablecaption{{\it Kepler} discoveries used for metallicity
accuracy analysis
\label{mettable}}
\tablehead{
\colhead{Name} & \colhead{KIC number} & \colhead {Spectroscopic [Fe/H]} &
\colhead{Reference} }
\startdata
TrES-2 host star & 11446443 & $-0.15 \pm 0.1$ & \citet{torres08} \\
Kepler-4 & 11853905 & $0.17 \pm 0.06$ & \citet{borucki10} \\
Kepler-7 & 5780885 & $0.11 \pm 0.03$ & \citet{latham10} \\
Kepler-8 & 6922244 & $-0.06 \pm 0.03$ & \citet{jenkins10} \\
Kepler-9 & 3323887 & $0.12 \pm 0.04$ & \citet{holman10} \\
Kepler-10 & 11904151 & $-0.15 \pm 0.04$ & \citet{batalha11} \\
Kepler-11 & 6541920 & $0.00 \pm 0.10$ & \citet{lissauer11} \\
Kepler-12 & 11804465 & $0.07 \pm 0.04$ & \citet{fortney11} \\
Kepler-14 & 10264660 & $0.12 \pm 0.06$ & \citet{buchhave11} \\
Kepler-15 & 11359879 & $0.36 \pm 0.07$ & \citet{endl11} \\
Kepler-16AB & 12644769 & $-0.3 \pm 0.2$ & \citet{doyle11} \\
Kepler-17 & 10619192 & $0.26 \pm 0.1$ & \citet{bonomo11} \\
Kepler-18 & 8644288 & $0.2 \pm 0.04$ & \citet{cochran11} \\
Kepler-19 & 2571238 & $-0.13 \pm 0.06$ & \citet{ballard11} \\
Kepler-20 & 6850504 & $0.02 \pm 0.04$ & \citet{fressin11} \\
Kepler-22 & 10593626 & $-0.29 \pm 0.06$ & \citet{borucki11b} \\
Kepler-23 & 11512246 & $-0.09 \pm 0.14$ & \citet{ford12} \\
Kepler-26 & 9757613 & $-0.21 \pm 0.08$ & \citet{steffen12} \\
Kepler-27 & 5792202 & $0.41 \pm 0.04$ & \citet{steffen12} \\
Kepler-30 & 3832474 & $0.18 \pm 0.027$ & \citet{fabrycky12} \\
Kepler-34AB & 8572936 & $-0.07 \pm 0.15$ & \citet{welsh12} \\
Kepler-35AB & 9837578 & $-0.34 \pm 0.2$ & \citet{welsh12} \\
KOI-135 & 9818381 & $0.33 \pm 0.11$ & \citet{bonomo11} \\
KOI-196 & 9410930 & $-0.09 \pm 0.16$ & \citet{santerne11} \\
KOI-204 & 9305831 & $0.26 \pm 0.1$ & \citet{bonomo11} \\
KOI-254	& 5794249 & $0.13 \pm 0.13$ & \citet{johnson11} \\
KOI-423 & 9478990 & $-0.29 \pm 0.1$ & \citet{bouchy11} \\
\nodata & 6185331 & $0.11 \pm 0.15$ & \citet{fischer12} \\
\enddata
\end{deluxetable*}

\end{document}